\documentclass[aps, prl, twocolumn, superscriptaddress, longbibliography, preprintnumbers, floatfix]{revtex4-2}

\usepackage{orcidlink}
\usepackage{graphicx}
\usepackage{xcolor}
\usepackage{amsmath,amsfonts,amsthm,amssymb}
\usepackage{mathtools}
\usepackage{bm}
\usepackage{verbatim}
\usepackage{multirow}
\usepackage{hyperref}

\newcommand{\orcid}[1]{\href{https://orcid.org/#1}{\includegraphics[width=10pt]{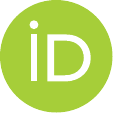}}}
\newcommand{\ror}[1]{\href{https://ror.org/#1}{\includegraphics[width=8pt]{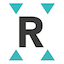}}}

\hypersetup{
    pdfnewwindow=true,      
    colorlinks=true,       
    linkcolor=violet,          
    citecolor=violet,        
    filecolor=violet,      
    urlcolor=violet        
}

\begin{document}

\title{Towards Measuring the CP-Violating Phase with Atmospheric Neutrinos}

\author{John F. Beacom \orcid{0000-0002-0005-2631}}
\email{beacom.7@osu.edu}
\affiliation{Center for Cosmology and AstroParticle Physics (CCAPP), \href{https://ror.org/00rs6vg23}{The Ohio State University}, Columbus, OH 43210, USA}
\affiliation{Department of Physics, \href{https://ror.org/00rs6vg23}{The Ohio State University}, Columbus, OH 43210, USA}
\affiliation{Department of Astronomy, \href{https://ror.org/00rs6vg23}{The Ohio State University}, Columbus, OH 43210, USA}

\author{Nicole F. Bell \orcidlink{0000-0002-5805-9828}}
\email{n.bell@unimelb.edu.au}
\affiliation{School of Physics, \href{https://ror.org/01ej9dk98}{The University of Melbourne}, Victoria 3010, Australia}
\affiliation{ARC Centre of Excellence for Dark Matter Particle Physics, School of Physics, \href{https://ror.org/01ej9dk98}
{The University of Melbourne}, Victoria 3010, Australia}

\author{Matthew J. Dolan~\orcidlink{0000-0003-3420-8718}}
\email{matthew.dolan@unimelb.edu.au}
\affiliation{School of Physics, \href{https://ror.org/01ej9dk98}{The University of Melbourne}, Victoria 3010, Australia}
\affiliation{ARC Centre of Excellence for Dark Matter Particle Physics, School of Physics, \href{https://ror.org/01ej9dk98}
{The University of Melbourne}, Victoria 3010, Australia}

\author{\\ Stephan A. Meighen-Berger \orcidlink{0000-0001-6579-2000}}
\email{stephan-meighen-berger@uiowa.edu}
\affiliation{School of Physics, \href{https://ror.org/01ej9dk98}{The University of Melbourne}, Victoria 3010, Australia}
\affiliation{\href{https://ror.org/036jqmy94}{University of Iowa}, Iowa City, Iowa 52242, USA}

\author{Ho Man Yim \orcidlink{0009-0000-7963-5313}}
\email{homan.yim@student.unimelb.edu.au}
\affiliation{School of Physics, \href{https://ror.org/01ej9dk98}{The University of Melbourne}, Victoria 3010, Australia}
\affiliation{ARC Centre of Excellence for Dark Matter Particle Physics, School of Physics, \href{https://ror.org/01ej9dk98}
{The University of Melbourne}, Victoria 3010, Australia}

\date{\today}


\begin{abstract}
We propose a new approach, differing significantly from prior work, to measure the CP-violating phase in neutrino mixing using atmospheric neutrinos. Our approach uses flavor-ratio observables for sub-GeV atmospheric neutrinos that incorporate realistic detection effects and reduce systematic uncertainties.  For the example of Hyper-Kamiokande --- the first experiment with sufficient atmospheric-neutrino statistics in this energy range --- \textit{our approach can surpass the sensitivity of the Tokai to Hyper-Kamiokande (T2HK) long-baseline experiment near $\mathit{\delta_\mathrm{CP} = 90^\circ}$ and $\mathit{270^\circ}$}.  Realizing this potential will require additional, but realistic, work to reduce theoretical uncertainties.  Success would provide an important complement to multi-\$1B accelerator-based measurements.
\end{abstract}

\maketitle


\textbf{\textit{Introduction.---}}
The observed dominance of matter over antimatter in the Universe is a profound mystery~\cite{Steigman:1976ev, Dine:2003ax}.  It has long been known that any solution to this mystery must include charge-parity (CP) violation, baryon or lepton number violation, and a departure from thermal equilibrium~\cite{Sakharov:1967dj}.  
As CP violation in the Standard-Model quark sector is insufficient for baryogenesis~\cite{Gavela:1993ts, Huet:1994jb}, it is essential to test CP violation in the neutrino sector, as it may enable leptogenesis~\cite{Fukugita:1986hr, Buchmuller:2005eh, Davidson:2008bu}. 

Leptonic CP violation may arise via a complex phase, $\delta_\mathrm{CP}$, in the Pontecorvo-Maki-Nakagawa-Sakata (PMNS) neutrino mixing matrix~\cite{Pontecorvo:1967fh, Pontecorvo:1957cp, Pontecorvo:1957qd, Maki:1962mu, ParticleDataGroup:2024cfk}. Measuring $\delta_\mathrm{CP}$ is a central goal of the worldwide program of accelerator long-baseline neutrino experiments, which directly probe CP-violating effects set by the Jarlskog invariant, $J \propto \sin\delta_\mathrm{CP}$~\cite{Jarlskog:1985cw}. Experiments to date have found encouraging but conflicting hints that $\delta_\mathrm{CP}$ is nonzero~\cite{T2K:2019bcf, NOvA:2021nfi, T2K:2023smv, T2K:2025wet}.  Definitive answers are expected from two forthcoming megaprojects:~Tokai to Hyper-Kamiokande (T2HK)~\cite{Hyper-Kamiokande:2018ofw, Hyper-Kamiokande:2025fci} and the Deep Underground Neutrino Experiment (DUNE)~\cite{DUNE:2020ypp, DUNE:2020jqi}.

Because $\delta_\mathrm{CP}$ is of such fundamental importance --- and is so hard to measure --- we need multiple \textit{independent} approaches to probe it. In principle, atmospheric neutrinos offer a complementary probe.  Several studies explored this~\cite{Freund:2001pn, Farzan:2002ct, Peres:2003wd, Peres:2009xe, Akhmedov:2012ah, Razzaque:2014vba, Kelly:2019itm, Martinez-Soler:2019nhb, Minakata:2019gyw, Ternes:2019sak, Indumathi:2021xtb, Arguelles:2022hrt, Super-Kamiokande:2023ahc, Chatterjee:2024ein, Barenboim:2024dye, Birkenfeld:2025cxe, ESSnuSB:2026dar}, with some finding few-$\sigma$ sensitivity to CP violation in the best cases.  The limitations mainly arise from difficulties in distinguishing neutrinos from antineutrinos and reconstructing neutrino kinematics. 

In this \textit{Letter}, we propose a new approach to measure $\delta_\mathrm{CP}$ with atmospheric neutrinos.  Instead of $\sin\delta_\mathrm{CP}$, we focus on  $\cos\delta_\mathrm{CP}$, which indirectly probes CP violation, building on the initial considerations of Refs.~\cite{Kimura:2007mu, Denton:2023qmd}. Measuring both $\cos\delta_\mathrm{CP}$ and $\sin\delta_\mathrm{CP}$ could break degeneracies and provide complementary sensitivity over the full $\delta_\mathrm{CP}$ range. \textit{To isolate $\mathit{\cos\delta_\mathrm{CP}}$, we develop a new observable based on the $\mathit{e/\mu}$ flavor ratio for sub-GeV atmospheric neutrinos that does not require separating neutrinos and antineutrinos.}   While we focus on Hyper-Kamiokande, the potential of our work is more general. In current full fits for Super-Kamiokande~\cite{Super-Kamiokande:2023ahc} or projected for Hyper-Kamiokande~\cite{Hyper-Kamiokande:2018ofw}, the sensitivity we predict to $\cos\delta_\mathrm{CP}$ is not yet evident. Key reasons are their conservative uncertainties on the flux flavor ratio and cross-section flavor ratio (both improving but needing more work) and the neutrino-mixing uncertainties (rapidly improving in independent measurements).

In the following, we review $\delta_\mathrm{CP}$ sensitivity with accelerator neutrinos, show the potential of atmospheric neutrinos, provide detailed calculations, then conclude.  In End Matter, we provide further details.


\begin{figure*}[t]
\centering
\includegraphics[width=0.95\textwidth]{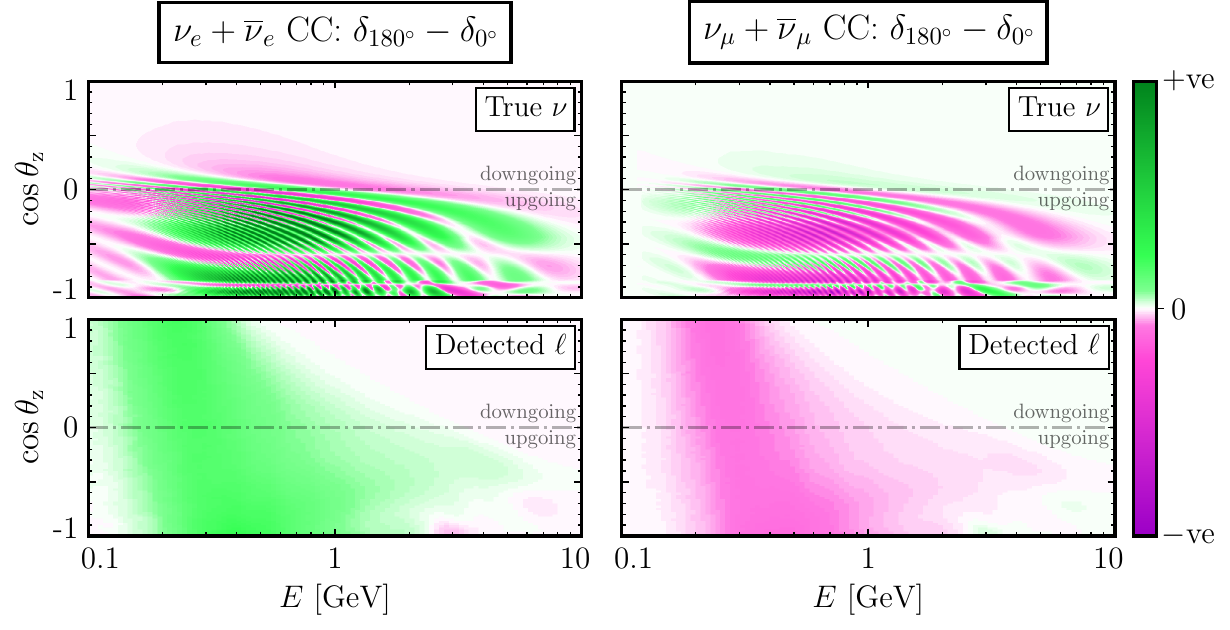}
\caption{Difference ($\delta_\mathrm{CP} = 180^\circ$ versus $\delta_\mathrm{CP} = 0^\circ$) in CC interaction rates for $\nu_e + \bar{\nu}_e$ (left) versus $\nu_\mu + \bar{\nu}_\mu$ (right). In the upper panels, the variables are the incoming neutrino energy and zenith angle; in the lower panels, they are the outgoing charged-lepton energy and zenith angle. \textit{Even including realistic detection effects, clear dependencies on $\mathit{\delta_\mathrm{CP}}$ remain.}}
\label{fig:flavor_oscillagram}
\end{figure*}

\textbf{\textit{Neutrino Mixing Framework.---}}
Neutrino mixing arises from the misalignment of the flavor ($\nu_\alpha$) and mass ($\nu_i$) eigenstates, as encoded in the PMNS matrix, $U$. In the standard parameterization~\cite{ParticleDataGroup:2024cfk}, this matrix has three mixing angles ($\theta_{12}$, $\theta_{13}$, $\theta_{23}$) and one nontrivial phase ($\delta_\mathrm{CP}$). In vacuum, the neutrino flavor transition probabilities, $P_{\alpha\beta} \equiv P_{\nu_\alpha\rightarrow \nu_\beta}$ ($\alpha,\beta = \{e,\mu,\tau\}$), are~\cite{Bilenky:1978nj, Giunti:1053706}:
\begin{equation}
    \begin{split}
    P_{\alpha \beta} = \delta_{\alpha \beta} 
    & -4 \sum_{k>j} \text{Re} \left[U_{\alpha k}^{*} U_{\beta k} U_{\alpha j} U_{\beta j}^{*}\right] \sin ^{2}\left(\Phi_{k j}\right) \\
    & +2 \sum_{k>j} \text{Im} \left[U_{\alpha k}^{*} U_{\beta k} U_{\alpha j} U_{\beta j}^{*}\right] \sin \left(2\Phi_{k j}\right),
    \end{split}
\label{eq:vacuum_probability}
\end{equation}
where the oscillation phase, $\Phi_{k j} \equiv \Delta m_{k j}^{2} L/4E_\nu$ ($k,j = 1,2,3$), depends on the neutrino mass splitting $\Delta m_{k j}^{2}$, baseline $L$, and energy $E_\nu$. The sums over mass eigenstates are split into real and imaginary parts, and $\delta_{\alpha \beta}$ is the Kronecker delta.  The real part of Eq.~(\ref{eq:vacuum_probability}) is CP-even, with all terms being proportional to 1, $\cos\delta_\mathrm{CP}$, or $\cos^2\delta_\mathrm{CP}$, whereas the imaginary part is CP-odd and proportional to $\sin\delta_\mathrm{CP}$.  The antineutrino flavor transition probabilities, $\overline{P_{\alpha \beta}}$, are obtained from Eq.~(\ref{eq:vacuum_probability}) via $\delta_\mathrm{CP} \rightarrow - \delta_\mathrm{CP}$.  We use the best-fit mixing angles and mass splittings from the global fits of Ref.~\cite{Esteban:2024eli} while taking $\delta_\mathrm{CP}$ to be unconstrained.  We assume the normal ordering, though this choice only slightly affects our results.


\textbf{\textit{Accelerator Neutrinos and $\boldsymbol{\delta_\mathrm{CP}}$.---}}
For T2HK, we estimate the $\delta_\mathrm{CP}$ sensitivity for later contrast with atmospheric neutrinos.  We take representative parameters of $E_\nu = 0.6$~GeV and $L = 295$~km~\cite{Hyper-Kamiokande:2025fci} and neglect matter effects. We find:
\begin{equation}
\begin{split}
    P_{\mu e} &\approx 0.04 \left(1 - 0.30\sin\delta_\mathrm{CP} + 0.01\cos\delta_\mathrm{CP} \right)\\
    \overline{P_{\mu e}} &\approx 0.04 \left(1  + 0.30\sin\delta_\mathrm{CP} + 0.01\cos\delta_\mathrm{CP}\right)\,.
\end{split}
\label{eq:T2HK-appearance}
\end{equation}
The $\sin\delta_\mathrm{CP}$ terms (prefactors proportional to $\sin\Phi_{32}$) are much larger than the $\cos\delta_\mathrm{CP}$ terms (proportional to $\cos\Phi_{32}$)~\cite{Nunokawa:2007qh, Banerjee:2026qng}.  This is because $E_\nu$ and $L$ were chosen to target the oscillation maximum, $\Phi_{32} \simeq 90^\circ$, which enhances the $\sin\delta_\mathrm{CP}$ terms but diminishes the $\cos\delta_\mathrm{CP}$ terms.  Furthermore, $\sin\delta_\mathrm{CP}$ can be isolated by comparing neutrino-antineutrino appearance at the same $L/E_\nu$, which cancels many uncertainties.  In contrast, there is no easy way to amplify and isolate the $\cos\delta_\mathrm{CP}$ terms with accelerator neutrinos. 


\textbf{\textit{Potential of Atmospheric Neutrinos.---}}
To build intuition, we start with \textit{difference oscillograms} that illustrate how atmospheric neutrino charged-current (CC) event rates for $\nu_e + \bar{\nu}_e$ versus $\nu_\mu + \bar{\nu}_\mu$ depend on $\delta_\mathrm{CP}$.  As detailed below, we include all relevant flux, mixing (including matter effects), interaction, and detection effects.

Figure~\ref{fig:flavor_oscillagram} shows how the CC rates for each flavor depend on $\delta_\mathrm{CP}$, for the extreme example of  $\delta_\mathrm{CP} = 180^\circ$ versus $\delta_\mathrm{CP} = 0^\circ$, which best isolates $\cos\delta_\mathrm{CP}$.  In the upper panels, the variables are the incoming neutrino energy and zenith angle.  Three important features are evident:
\begin{enumerate}

    \item For upgoing neutrinos, the fine oscillatory structures arise from the atmospheric mass splitting, while the broad oscillatory structures arise from the solar mass splitting.  There is a clear dependence on $\delta_\mathrm{CP}$, revealed by the intensity of the color.
    
    \item For downgoing neutrinos, where solar-driven oscillations only partially develop due to the short distances, the overall oscillations approach the two-flavor limit, which is intrinsically $\delta_\mathrm{CP}$-independent.

    \item For this example, the differences due to $\delta_\mathrm{CP}$ variation are mostly of one sign, with the differences for $\nu_e + \bar{\nu}_e$ and $\nu_\mu + \bar{\nu}_\mu$ having opposite signs, indicated by their opposite colors. 
    
\end{enumerate}

In the lower panels, the variables are instead the outgoing charged-lepton energy and zenith angle, showing what could be detected in Hyper-Kamiokande, including all relevant effects.  The oscillation patterns are smeared out, reducing sensitivity to the mixing parameters and the mass ordering.  \textit{However, sensitivity to $\mathit{\delta_\mathrm{CP}}$ remains and is maximized in the sub-GeV range~\cite{Freund:2001pn, Razzaque:2014vba, Minakata:2019gyw, Kelly:2019itm, Martinez-Soler:2019nhb}.}  At lower energies, the decreasing flux and cross-section suppress the sensitivity; at higher energies, matter effects suppress solar mixing, driving oscillations towards the two-flavor limit. 


\textbf{\textit{Analytical Estimates.---}}
The dependence of atmospheric neutrino oscillations on $\delta_\mathrm{CP}$ was widely known.  Prior work exploited this, but found only few-$\sigma$ sensitivity to CP violation even for optimistic detection scenarios.  Most focused on probing $\sin\delta_\mathrm{CP}$, which has two challenges.  First, there must be a way to separate neutrinos from antineutrinos.  Second, there must be a way to resolve neutrino oscillation structures even after detection smearing effects are included.

In contrast, to isolate $\cos\delta_\mathrm{CP}$, we take a simple and robust approach using the $e/\mu$ \textit{flavor ratio},
\begin{equation}
    R_{e/\mu} = \Gamma_\text{e}/\Gamma_{\mu}\,,
\end{equation}
where $\Gamma$ is an event rate integrated over \textit{all} zenith angles and a wide range of detected energies, and including both neutrinos and antineutrinos.  As suggested in Fig.~\ref{fig:flavor_oscillagram}, the power of this ratio arises from the different $\delta_\mathrm{CP}$ dependence of $\nu_e + \bar{\nu}_e$ versus $\nu_\mu + \bar{\nu}_\mu$ oscillations.

We begin by making analytical estimates for $R_{e/\mu}$ analogous to Eq.~(\ref{eq:T2HK-appearance}).  For simplicity, we only consider neutrinos in these estimates, while our full calculation below includes both $\nu$ and $\overline\nu$.  Further, we temporarily neglect the energy and angular dependence of fluxes and cross-sections, and approximate the initial flavor ratio as $\phi_e:\phi_\mu:\phi_\tau = 1:2:0$~\cite{Honda:2016qyr, Super-Kamiokande:2023ahc, Kashti:2005qa, Kajita:2016cak}.  

For upgoing neutrinos, where the dependence on $\delta_\mathrm{CP}$ originates, the interaction rates scale as 
\begin{equation}
	\Gamma_{\alpha}^\text{up} 
    \propto \int^0_{-1} d\cos\theta_z\, \int dE_\nu\  \left(P_{e\alpha} + 2 P_{\mu \alpha}\right),
\label{eq:Upgoing rate}
\end{equation}
where $\alpha =e, \mu$.  Two effects are important.  First, these neutrinos travel through Earth and experience significant matter effects.   At sub-GeV energies and typical Earth densities, this primarily modifies $\theta_{12}$ and slightly modifies $\theta_{13}$.  Following Refs.~\cite{Denton:2016wmg, Denton:2018hal, Xing:2018lob, Ioannisian:2018qwl, Denton:2023qmd}, we use 
\begin{equation}
    \cos2\theta_{12}\rightarrow \Big(\cos2\theta_{12} - a \cos^2 \theta_{13}/\Delta m^2_{21}\Big)/S_\odot\,,
\label{eq:modified solar mixing angle}
\end{equation}
where $a = 2\sqrt2 G_F N_e E_\nu$ is the matter potential and 
\begin{equation}
    S_{\odot} = \sqrt{(\cos2\theta_{12} - a \cos^2 \theta_{13}/\Delta m^2_{21})^2+\sin^22\theta_{12}} \,.
\label{eq:Solar modification}
\end{equation}
For a typical neutrino energy of $E_\nu =0.5$~GeV and a mantle-like density of 5~g/cm$^3$, $S_{\odot} = 2.28$.  

Second, the broad range of energies and travel distances averages the oscillations.  For this simple estimate, we approximate $\sin\left(2\Phi\right) \rightarrow 0$ and $\sin^{2}\left(\Phi\right) \rightarrow 1/2$ for all $\Phi_{ij}$ in the analog of Eq.~(\ref{eq:vacuum_probability}) that uses matter-modified mixing angles.  Importantly, this averaging removes the $\sin\delta_\mathrm{CP}$ dependence but retains the $\cos\delta_\mathrm{CP}$ terms.  It also reduces sensitivity to the uncertainties on the mass splittings and mass ordering~\cite{Akhmedov:2012ah, Indumathi:2017kxa, Ioannisian:2020isl}.  

With the two effects above, the transition probabilities for upgoing neutrinos are $P_{ee} \simeq 0.884$,  $P_{\mu e} = P_{e \mu} \simeq (0.057 - 0.026\cos\delta_\mathrm{CP})$, and $P_{\mu \mu} \simeq (0.469 + 0.028\cos\delta_\mathrm{CP})$. Downgoing neutrinos travel distances short enough that their oscillations can be neglected, so $P_{ee} = P_{\mu\mu} \simeq 1 $ and $P_{\mu e} = P_{e \mu} \simeq 0$. Assuming comparable upgoing and downgoing fluxes, we estimate the total rates as
\begin{alignat}{2}
    \Gamma_{e} = \Gamma_{e}^\text{up}  + \Gamma_{e}^\text{down} & \propto (P_{ee}^\text{up} + 2 P_{\mu e}^\text{up}) + P_{ee}^\text{down}  \notag\\
    &\propto (0.998 - 0.052\cos\delta_\mathrm{CP})  + 1 \,,
    \label{eq:analytic_rates_e}
     \\
     \notag\\
    \Gamma_{\mu} = \Gamma_{\mu}^\text{up}  + \Gamma_{\mu}^\text{down} & \propto (P_{e\mu}^\text{up} + 2 P_{\mu \mu}^\text{up}) + 2P_{\mu \mu}^\text{down}  \notag\\
     &\propto (0.995 + 0.030\cos\delta_\mathrm{CP}) + 2 \,.
    \label{eq:analytic_rates_mu}
\end{alignat}
Importantly, because the electron and muon rates depend on $\cos\delta_\mathrm{CP}$ with opposite signs, taking their ratio enhances the effect: 
\begin{equation}
    R_{e/\mu} \propto  \frac{2(1 - 0.03\cos\delta_\mathrm{CP})}{3(1 + 0.01\cos\delta_\mathrm{CP})}\simeq 0.67(1 - 0.04\cos\delta_\mathrm{CP}) \,. 
\label{eq:analytic flavor ratio}
\end{equation}
This ratio cancels overall uncertainties in the flux and cross-section, leaving those on the ratios between flavors.  While the estimated modulation is small ($4\%$), it would potentially be measurable with Hyper-Kamiokande, which will collect $\mathcal{O}$(100k) sub-GeV events in a ten-year exposure.


\textbf{\textit{Simulation of Atmospheric Events.---}}
Before presenting our full numerical results, we summarize how we create a realistic simulation of atmospheric neutrino events in Hyper-Kamiokande, a water-Cherenkov detector with a fiducial volume of 186~kton now under construction in Japan~\cite{Hyper-Kamiokande:2018ofw}.  The simulation techniques we use are based on our earlier work for Super-Kamiokande, Hyper-Kamiokande, and JUNO~\cite{Beacom:2003zu, Bell:2020rkw, Bell:2021esh, Suliga:2023pve, Meighen-Berger:2023xpr, Zhou:2023mou, Meighen-Berger:2025pcq}.

We define our observables in terms of final state charged leptons from CC interactions. We calculate the expected spectra as 
\begin{equation}
    \frac{\mathrm{d}N_{\ell_\alpha}}{\mathrm{d}E_\mathrm{det}} = 
    \frac{\mathrm{d}\phi_{\nu_\alpha}}{\mathrm{d}E_\nu} \otimes
    \mathcal{C}_{\alpha}(E_\nu, E_\mathrm{det})\otimes
    \epsilon_{\alpha}(E_\mathrm{det}),
    \label{eq:rates}
\end{equation}
where $\alpha \in \{e, \mu\}$ denotes the flavor of lepton and neutrino.  (Our calculations are also differential in the angles and take into account the possibility of charged-lepton mis-ID, but we suppress the notation.)  There are three terms in the convolution:

The first term is the oscillated neutrino spectra. For the initial unoscillated atmospheric neutrino fluxes, we use the site- and solar-phase-dependent atmospheric neutrino fluxes from HKKM~\cite{Honda:2016qyr, HKKM_Data}.  We calculate the three-flavor neutrino oscillations in Earth with full matter effects using {\tt nuCraft} and assuming the Preliminary Earth Reference Model (PREM)~\cite{Dziewonski:1981xy, Wallraff:2014qka, nucraft}.  

The second term encodes the mapping between neutrinos and detected charged leptons, including detector resolution effects.  The lepton distribution from primary neutrino interactions is obtained from {\tt GENIE 3.2.0} with tune G18\_10a\_02\_11b~\cite{Andreopoulos:2009rq, Andreopoulos:2015wxa, GENIE:2021zuu}. We generate $10^7$ neutrino interactions in water with $E_\nu$ in the range 0.2--10~GeV. We use an injection spectrum scaling as $1/E_\nu$ to sample uniformly in $\log E_\nu$, later reweighting this to match the oscillated atmospheric spectrum. We then simulate final-state particle propagation using {\tt GEANT4} to incorporate detector effects~\cite{GEANT4:2002zbu}.  We require that the final states have one charged lepton and no other relevant particles (pions, etc.), and that all energy is deposited within the fiducial volume.  These fully contained single-ring events are dominated by charged-current quasielastic interactions~\cite{Formaggio:2013kya, Alvarez-Ruso:2017oui}.  The energies of the charged leptons are estimated by their Cherenkov light yields, approximately 6 per event per MeV (photoelectrons/MeV), about the same as for Super-Kamiokande~\cite{Hyper-Kamiokande:2018ofw, Suzuki:2019jby, ParticleDataGroup:2024cfk}.

The third term is the detection efficiency, for which we adopt results from Super-Kamiokande, $\epsilon_e \geq 85\%$ and $\epsilon_\mu \geq 90\%$~\cite{Super-Kamiokande:2019gzr}.

We validate our energy spectra and angular distributions via our similar codes for Super-Kamiokande, finding good agreement with their data~\cite{Super-Kamiokande:2005mbp, Super-Kamiokande:2010orq, Super-Kamiokande:2017yvm, Super-Kamiokande:2019gzr, Super-Kamiokande:2023ahc}. 


\textbf{\textit{Atmospheric-Neutrino Sensitivity to $\boldsymbol{\delta_\mathrm{CP}}$.---}}
We now define our analysis region for calculating $R_{e/\mu}$ in terms of detected charged-lepton energy and angle.  For both $e$- and $\mu$-like events (now containing both neutrino and antineutrino contributions), we include \textit{all} zenith angles but restrict the detected energies to the range 0.2--1.3~GeV (Super-Kamiokande's sub-GeV range). This energy range is chosen so that (i) the $\delta_\mathrm{CP}$-dependence is maximized (see Fig.~\ref{fig:flavor_oscillagram}), (ii) neutrino interactions are dominant over antineutrino interactions~\cite{Formaggio:2013kya, Andreopoulos:2009rq, Zhou:2023mou}, increasing sensitivity to $\delta_\mathrm{CP}$, and (iii) quasielastic interactions dominate, leading to good event identification and reasonably well-understood kinematics~\cite{Super-Kamiokande:2019gzr, Suzuki:2019jby, Zhou:2023mou}.

For a ten-year exposure of Hyper-Kamiokande, we expect $5.5 \times 10^4$ fully contained $e$-like charged-current events and $7.2 \times 10^4$ fully contained $\mu$-like charged-current events for the sub-GeV range.

\begin{figure}[t]
\centering
\includegraphics[width=0.99\columnwidth]{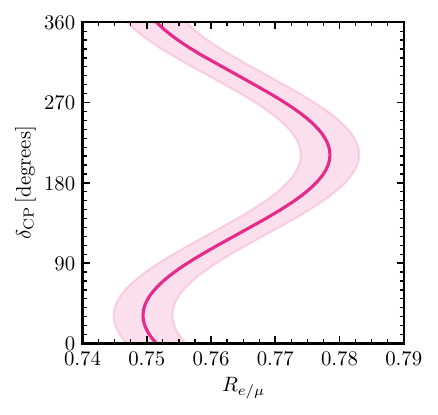}
\caption{Expected $R_{e/\mu}$ ratio from our full simulation result, as given by Eq.~(\ref{eq:simulation_flavor_ratio}).  The axes are interchanged to illustrate how a measurement of the flavor ratio would constrain $\delta_\mathrm{CP}$.} 
\label{fig:flavor_ratio}
\end{figure}

One of our main results is 
\begin{equation}
    R_{e/\mu} \simeq  0.760(1 - 0.017\cos\delta_\mathrm{CP} - 0.009\sin\delta_\mathrm{CP}),
\label{eq:simulation_flavor_ratio}
\end{equation}
which we obtain from our full simulation (and illustrate in Fig.~\ref{fig:flavor_ratio} in End Matter).  Taking into account both terms, the effective modulation is $\simeq2\%$.  This is the realistic version of our analytical estimate in Eq.~(\ref{eq:analytic flavor ratio}).  While largely in agreement, there are three important differences.  First, here the overall prefactor is shifted upwards because we now correctly account for the oscillations of downgoing neutrinos.  Second, the prefactor for $\cos\delta_\mathrm{CP}$ is reduced. This is mainly due to the inclusion of antineutrinos, for which the prefactors of $\cos\delta_\mathrm{CP}$ have opposite signs to neutrinos, though neutrinos dominate the yields (see End Matter).  Third, there is a subdominant $\sin\delta_\mathrm{CP}$ component that arises because the oscillations are not completely averaged. 

To estimate the sensitivity to $\delta_\mathrm{CP}$, we include statistical and systematic uncertainties.  An important example of the latter is on the unoscillated $\nu_e/\nu_\mu$ flux ratio. In Super-Kamiokande atmospheric flux analyses, this uncertainty is set as large as $\sim$3\%~\cite{Super-Kamiokande:2015qek} in the neutrino energy ranges of interest.  However, it is expected to be below 1\%~\cite{Barr:2006it} or below 2\%~\cite{Honda:2006qj, Honda:2015fha}.  For this and the closely related uncertainty on the $\nu_e/\nu_\mu$ cross-section ratio, we assume that it will be possible to reach 1\% with further work, discussed below.  We also include uncertainties on the cross-section normalization ($16\%$), the up-down flux asymmetry ($2\%$), and the
neutral-current background normalization ($10\%$).  Last, we include uncertainties on the Earth density profile and the mixing parameters, which are subdominant (see End Matter). 

\begin{figure}[t]
\centering
\includegraphics[width=0.99\columnwidth]{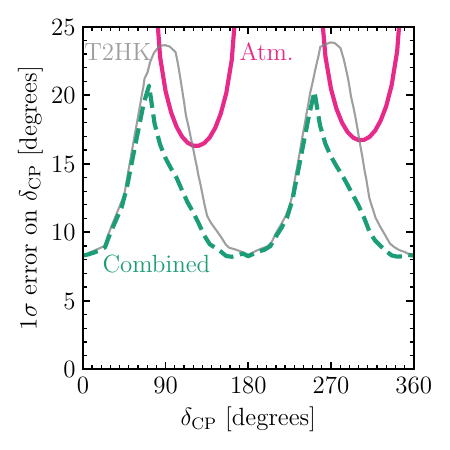}
\caption{Expected 1$\sigma$ resolution on $\delta_\mathrm{CP}$ as a function of its true value.  We show results for atmospheric neutrinos (this work), T2HK accelerator neutrinos~\cite{Hyper-Kamiokande:2025fci}, and their combination. The competitive complementarity is evident.}
\label{fig:sensitivity}
\end{figure}

Figure~\ref{fig:sensitivity} shows the projected 1$\sigma$ resolution on $\delta_{\rm{CP}}$ that can be attained with our approach and ten years of sub-GeV atmospheric neutrinos in Hyper-Kamiokande.  We compute the predicted values of $R_{e/\mu}$ over the full range of $\delta_\mathrm{CP}$ and construct likelihoods to determine the expected precision.  When uncertainties are asymmetric, we conservatively choose the larger one.  The best uncertainty obtained on $\delta_\mathrm{CP}$ is $\simeq$$17^\circ$, which is achieved near but not quite at $\delta_\mathrm{CP} = 90^\circ$ and $270^\circ$. The reason for this shift is the $\sin\delta_\mathrm{CP}$ term shown in Eq.~(\ref{eq:simulation_flavor_ratio}). The depths of the two minima are slightly different due to a small $\cos^2 \delta_\mathrm{CP}$ term, not shown in Eq.~(\ref{eq:simulation_flavor_ratio}). At other angles, the precision worsens but never exceeds $\simeq$$70^\circ$, which would still be a useful independent result.  If we relax our assumed flavor-ratio uncertainty from 1\% to 2\% at Hyper-Kamiokande statistics, the best precision on $\delta_\mathrm{CP}$ worsens from $\sim$$17^\circ$ to $\sim$$32^\circ$.  If we assume inverted instead of normal ordering, the results are worse by just a few degrees.

Figure~\ref{fig:sensitivity} also shows the projected 1$\sigma$ resolution for the T2HK accelerator experiment with ten years of exposure and Super-Kamiokande systematics~\cite{Hyper-Kamiokande:2025fci}.  Because T2HK primarily probes $\sin\delta_\mathrm{CP}$ rather than $\cos\delta_\mathrm{CP}$, its best precision is near $\delta_{\rm{CP}}=0^\circ$ and $180^\circ$, where it achieves a superior resolution of about  $8^\circ$.  But near maximal CP violation --- as suggested by current data~\cite{T2K:2019bcf, NOvA:2021nfi, T2K:2023smv, T2K:2025wet} --- \textit{our atmospheric approach achieves better precision than T2HK}.  Finally, Fig.~\ref{fig:sensitivity} shows the combined sensitivity of the accelerator and atmospheric approaches, illustrating the importance of a joint fit.  \textit{Combining the two approaches can also help resolve the two-fold degeneracy inherent in separate measurements of $\mathit{\sin\delta_\mathrm{CP}}$ and $\mathit{\cos\delta_\mathrm{CP}}$.}


\textbf{\textit{Conclusions and Next Steps.---}}
Do neutrinos violate CP symmetry and might this account for the observed matter-antimatter asymmetry of the universe?  Multi-\$1B accelerator long-baseline experiments are being built to find out.  If independent approaches can reach comparable sensitivity, that would be invaluable.  While the potential of atmospheric-neutrino measurements has long been recognized, the proposed approaches have not had adequate sensitivity~\cite{Freund:2001pn, Farzan:2002ct, Peres:2003wd, Peres:2009xe, Akhmedov:2012ah, Razzaque:2014vba, Kelly:2019itm, Martinez-Soler:2019nhb, Minakata:2019gyw, Ternes:2019sak, Indumathi:2021xtb, Arguelles:2022hrt, Super-Kamiokande:2023ahc, Chatterjee:2024ein, Barenboim:2024dye, Birkenfeld:2025cxe, ESSnuSB:2026dar}.

In this Letter, we demonstrate the first approach to measuring $\delta_{\rm{CP}}$ with atmospheric neutrinos that could exceed the sensitivity expected for the T2HK accelerator-neutrino project.  To be precise, this claim is only true near $\delta_{\rm{CP}} = 90^\circ$ and $270^\circ$, where CP is maximally violated, but that may be quite important.  The relevant atmospheric-neutrino data will be taken by Hyper-Kamiokande at the same time as it collects accelerator-neutrino data in the T2HK program. \textit{With both accelerator and atmospheric analyses, we will get the most robust results in the shortest time.}

To realize the potential of our approach, new work is needed to reduce theoretical uncertainties in the sub-GeV range.  In particular, the uncertainty on the initial $\nu_e/\nu_\mu$ flux ratio must be reduced to $1\%$, which can be done via studies of $\pi$-production in the atmosphere.  Similarly, reductions in uncertainties on the cross-section ratio are needed, which can be done in accelerator experiments. Given the high stakes of an independent $\delta_{\rm{CP}}$ measurement, the motivations to reduce these uncertainties will be high.  In future work, we will address these points, optimize the analysis, and extend it to other detection channels and experiments.

New analyses of Super-Kamiokande data could be an important addition to the global effort to probe $\delta_\mathrm{CP}$.  In a first, oversimplified analysis of data from Ref.~\cite{Super-Kamiokande:2023ahc}, corresponding to an exposure of 395~kton-years ($\sim$2 years of Hyper-K), we find a best fit of $\delta_\mathrm{CP} \sim (290 \pm 50)^\circ$ and that $\delta_\mathrm{CP} = 0^\circ$ is disfavored at $\sim$$1.8\sigma$.  This is nominally comparable to the accelerator result of Ref.~\cite{T2K:2024wfn}.  In agreement with Super-Kamiokande's atmospheric analysis, an uncertainty of $\gtrsim$2\% on the $e/\mu$ flavor ratio would remove their sensitivity to $\delta_\mathrm{CP}$ due to limited statistics. Last, we find that the sensitivity would be improved if neutron tagging with gadolinium were used to separate neutrinos versus antineutrinos~\cite{Beacom:2003nk, Super-Kamiokande:2024kcb}, which motivates running Super-Kamiokande with gadolinium longer.


\bigskip

\textbf{\textit{Acknowledgments.---}}
We are grateful for helpful discussions with Ivan Esteban, Matheus Hostert, Kevin Kelly, and Bei Zhou.  And we are especially grateful for helpful discussions with members of the Super-Kamiokande Collaboration (Thomas Leplumey, Ferey Mathieu, Benjamin Quilain, and Linyan Wan) that led to key improvements to our paper (see End Matter).

We used ChatGPT (OpenAI) and Claude (Anthropic) solely to assist with code development and to check for grammatical errors in the manuscript, which was written by us.  All AI-assisted code was reviewed and verified by the authors, who take full responsibility for the content of this manuscript.

This work was supported by the Australian Research Council through Discovery Project DP220101727 plus the University of Melbourne’s Research Computing Services and the Petascale Campus Initiative.  J.F.B. was supported by US National Science Foundation Grant No.\ PHY-2609870. J.F.B. also acknowledges visitor support from the World Research Hub (WRH) Program of Institute of Science Tokyo, the Sir Thomas Lyle Fellowship at the University of Melbourne, and the Institut Henri Poincaré (UAR 839 CNRS-Sorbonne Université) and LabEx CARMIN (ANR-10-LABX-59-01). M.J.D. also acknowledges visitor support from the CERN Theory Division. H.M.Y. was supported by a University of Melbourne Research Scholarship.


\clearpage

\onecolumngrid
\begin{center}
\textbf{\Large End Matter}   
\end{center}
\vspace*{0.2cm}
\twocolumngrid

Here we provide additional details that are helpful for reproducing our results, but which are not essential for the main arguments.  

\medskip


\textbf{\textit{Differences between arXiv v1/v2 and v3.---}}
This revised version (arXiv v3) of our paper has important differences from the original version (arXiv v1/v2). The bottom-line sensitivity is comparable even though our approach is quite different.

How is this possible? Because we had an error of interpretation, not an error of calculation.  In both versions of our paper, our final results were obtained with a full numerical treatment based on likelihoods calculated over the selected energy and angular ranges. Also in both versions, we sought to make the results easier to understand via an observable that captured the dominant contribution to the sensitivity.  In arXiv v1/v2, this was an up-down ratio, which was not the key driver.  In arXiv v3, we now realize that the key driver is the $e/\mu$ flavor ratio.  Our results are largely unchanged because we had implicitly included this in our full numerical analysis in v1/v2.

Discussions with members of the Super-Kamiokande Collaboration were critical for identifying these issues. In that process, we also found that their analyses adopt a cautious uncertainty on the $e/\mu$ flavor ratio and do not fully include external constraints on the mixing parameters. Separately, we corrected a bug in our implementation of the angular smearing, which led to modest changes in our final sensitivities.

\medskip


\textbf{\textit{Illustration of flavor ratio modulation.---}}
Figure~\ref{fig:flavor_ratio} shows how the flavor ratio changes with $\delta_\mathrm{CP}$, given in Eq.~(\ref{eq:simulation_flavor_ratio}).  The axes are flipped to illustrate how this figure would be used in practice to infer $\delta_\mathrm{CP}$ from a measured flavor ratio.  That equation can also be rewritten as $R_{e/\mu} \simeq 0.76 (1 - 0.02 \cos(\delta_\mathrm{CP}-27.9^\circ))$, where the sine term has here been absorbed as a phase shift of the cosine term.  Figure~\ref{fig:flavor_ratio} also shows the uncertainty band due to statistical and systematic uncertainties (see main text).  This replaces a figure in the arXiv v1/v2 version that was based on an up-down ratio.

\medskip


\textbf{\textit{Scaling of $R_{e/\mu}$ for Atmospheric $\nu + \bar{\nu}$.---}}
In the main text, we estimated the $\delta_\mathrm{CP}$ dependence of $R_{e/\mu}$ for neutrinos only.  Here, we estimate the scaling for antineutrinos.  As in the main text, we incorporate matter effects via Eqs.~(\ref{eq:modified solar mixing angle}) and (\ref{eq:Solar modification}). The matter potential changes sign for antineutrinos ($a \rightarrow -a$), so for $E_\nu =0.5$~GeV and a density of 5~g/cm$^3$, we obtain $S_{\odot} = 2.28 \rightarrow 2.99$.  The transition probabilities for upgoing antineutrinos are then $\overline{P_{ee}} \simeq 0.909$, $\overline{P_{\mu e}} =  \overline{P_{e \mu}} \simeq(0.046 + 0.022\cos\delta_\mathrm{CP})$, and $\overline{P_{\mu\mu}} \simeq (0.487 - 0.019 \cos\delta_\mathrm{CP})$.

To estimate the total rates, we again neglect the oscillations of downgoing antineutrinos and assume similar upgoing and downgoing fluxes, so that
\begin{alignat}{2}
    \overline{\Gamma_{e}} = \overline{\Gamma_{e}^\text{up}}  + \overline{\Gamma_{e}^\text{down}} & \propto (\overline{P_{ee}^\text{up}} + 2 \,\overline{P_{\mu e}^\text{up}}) + \overline{P_{ee}^\text{down}}  \notag\\
    &\propto (1.001 + 0.044\cos\delta_\mathrm{CP})  + 1 \,,
    \label{eq:antineutrino_analytic_rates_e}
     \\
     \notag\\
    \overline{\Gamma_{\mu}} = \overline{\Gamma_{\mu}^\text{up}}  + \overline{\Gamma_{\mu}^\text{down}} & \propto (\overline{P_{e\mu}^\text{up}} + 2 \, \overline{P_{\mu \mu}^\text{up}}) + 2 \,\overline{P_{\mu \mu}^\text{down}}  \notag\\
     &\propto (1.020 - 0.016\cos\delta_\mathrm{CP}) + 2 \,.
    \label{eq:antineutrino_analytic_rates_mu}
\end{alignat}
Note that the antineutrino and neutrino rates have opposite sign dependence on $\cos \delta_\mathrm{CP}$.  This could enhance sensitivity to $\cos\delta_\mathrm{CP}$ and help reduce systematics if the two samples could be distinguished.

To combine the neutrino rates from Eqs.~(\ref{eq:analytic_rates_e}, \ref{eq:analytic_rates_mu}) and the antineutrino rates from  Eqs.~(\ref{eq:antineutrino_analytic_rates_e}, \ref{eq:antineutrino_analytic_rates_mu}), we note that the antineutrino cross-section is $\simeq$3 times smaller at sub-GeV energies. This reduces the cancellation between the opposing $\cos\delta_\mathrm{CP}$ terms:
\begin{align}
    \text{e-like rate}
    & \simeq \Gamma_e + \frac{1}{3}\overline{\Gamma}_e
    \propto (2.665 - 0.037\cos\delta_\mathrm{CP})
    \label{eq:e-like-rate}\\[1em]
    \text{$\mu$-like rate}
    &= \Gamma_\mu + \frac{1}{3}\overline{\Gamma}_\mu \propto (4.002 + 0.025\cos\delta_\mathrm{CP}).
    \label{eq:mu-like-rate}
\end{align}
The $e/\mu$ ratio is then 
\begin{equation}
    R_{e/\mu} \propto \frac{2.665(1-0.014\cos\delta_\mathrm{CP})}{4.002(1+0.006\cos\delta_\mathrm{CP})} \simeq 0.67 (1-0.02\cos\delta_\mathrm{CP}),
\end{equation}
which is close to our full numerical result in Eq.~(\ref{eq:simulation_flavor_ratio}).

\medskip

\begin{figure}[t]
\centering
\includegraphics[width=0.99\columnwidth]{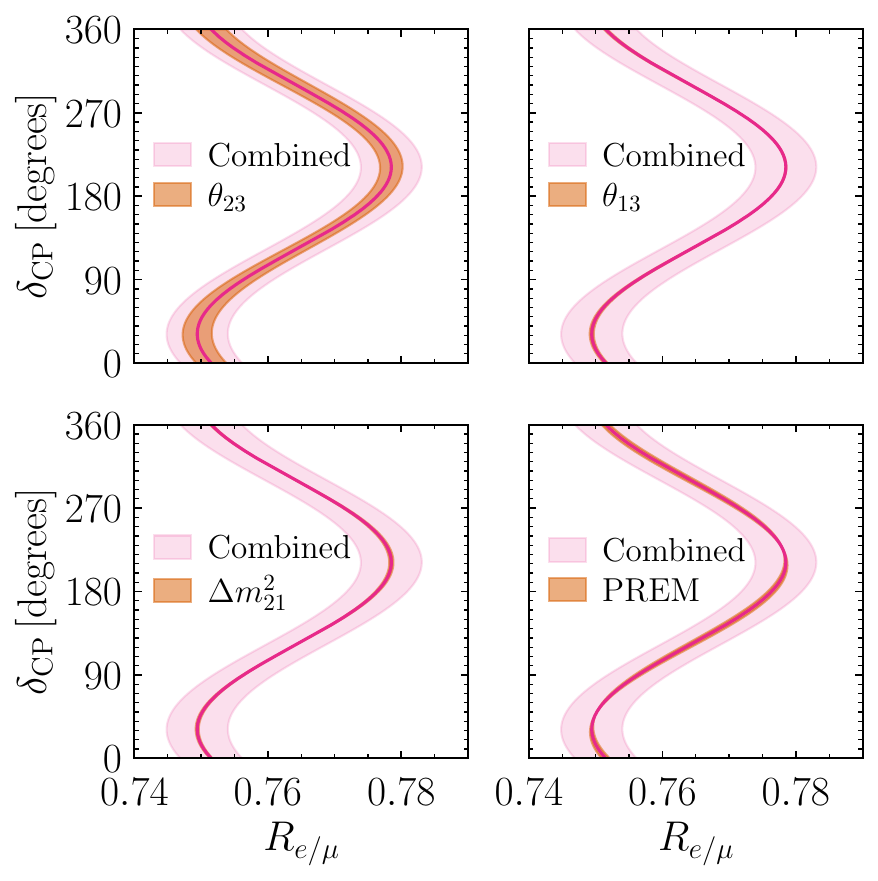}
\caption{Same as Figure~\ref{fig:flavor_ratio}, but now showing the labeled subdominant 1$\sigma$ uncertainties compared to the total.}
\label{fig:ratio_uncertainties}
\end{figure}

\textbf{\textit{Subdominant Uncertainties.---}}
Figure~\ref{fig:ratio_uncertainties} shows subdominant uncertainties due to Earth's density profile and mixing parameters (both included in our full numerical results) compared to the total uncertainty for $R_{e/\mu}$.

\begin{figure}[t]
\centering
\includegraphics[width=0.99\columnwidth]{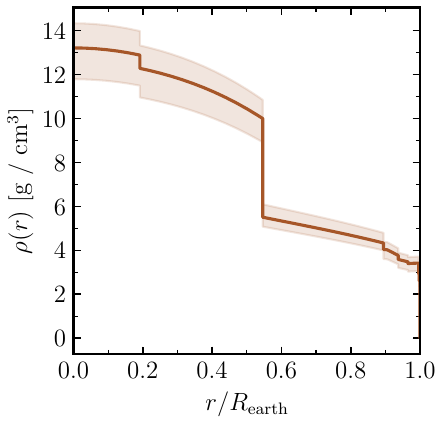}
\caption{Range of density profiles considered in assessing their impact on $R_{e/\mu}$. These are variations of the PREM model, constrained to keep the total Earth mass within $\pm 0.01\%$ of its nominal value.}
\label{fig:prem_variations}
\end{figure}

Figure~\ref{fig:prem_variations} shows the range of density profiles we used to assess their impact on $R_{e/\mu}$.  These profiles were generated based on PREM~\cite{Dziewonski:1981xy}, with the density of each layer (e.g., the mantle, outer core, inner core, and related structures) varied by up to $10\%$, while keeping the total mass within $\pm 0.01\%$ of the nominal Earth mass.  This constraint is conservative, given that CODATA restricts the variation on the total mass to $0.001\%$~\cite{Mohr:2015ccw}.

We assess the uncertainties in the mixing parameters by recomputing $R_{e/\mu}$ while varying each parameter within its 1$\sigma$ range from NuFit~\cite{Esteban:2024eli}.  The dominant uncertainties arise from $\theta_{23}$, $\theta_{13}$ and $\Delta m^{2}_{21}$ (shown in Fig.~\ref{fig:ratio_uncertainties}), as they control the leading oscillation amplitudes.  The last appears because the slower oscillations are not fully averaged out.  The remaining parameters have negligible effect on the ratio and are not shown in Fig.~\ref{fig:ratio_uncertainties}. Further, in the near future, the uncertainties on the mixing parameters will be significantly reduced~\cite{JUNO:2015zny, JUNO:2022mxj, Hyper-Kamiokande:2018ofw, Hyper-Kamiokande:2025fci}. 

\clearpage
\bibliography{bibliography}


\end{document}